\documentstyle[preprint,aps]{revtex}
\begin{document}
\title { Proton Spin in Chiral Quark Models }
\author{H. J. Weber,\quad X. Song }
\address{ Institute of Nuclear and Particle Physics, University of
Virginia, \\Charlottesville, VA 22901, USA}
\author{and}
\author{ M. Kirchbach}
\address{ Institut f\"ur Kernphysik, J. Gutenberg Universit\"at Mainz,
D--55099 Mainz, Germany}
\maketitle
\begin{abstract}
The spin and flavor fractions of constituent quarks in the proton are 
obtained from their chiral fluctuations involving Goldstone bosons. SU(3) 
breaking suggested by the mass difference between the strange and up, down 
quarks is included, and this improves the agreement with the data markedly.   
\end{abstract}
\newpage
\section{Introduction }
The nonrelativistic quark model (NQM) explains many of the properties 
of the nucleon and its excited states as originating from three
valence quarks whose dynamics is motivated by quantum chromodynamics (QCD), 
the gauge field theory of the strong interaction. The effective 
degrees of freedom at low energies are dressed or constituent quarks which are
expected to emerge in the spontaneous chiral symmetry breakdown of QCD 
that may be described by Nambu--Jona-Lasinio 
models~\cite{NJL}. The light quarks of QCD become dynamical quarks with mass 
$m_q(p^2)$ in this process. Upon approximating the dynamical mass by 
$m_q(0)\approx m_N/3$ one can introduce the concept of a constituent quark (of 
the NQM) at low momentum $p$. Along with dynamical quarks Goldstone 
bosons~\cite{GSW} occur as effective degrees of freedom in QCD below the 
chiral symmetry scale $4\pi f_\pi \approx 1169$ MeV for 
$f_\pi = 93$ MeV. 
Other degrees of freedom, such as gluons, are integrated out.
\par  
Chiral quark models which include these effective degrees of freedom have been 
developed for a long time starting with the Gell-Mann--Levy $\sigma$ 
model~\cite{GML}. The nonlinear $\sigma$ model is a 
starting point for soliton or Skyrme models of the nucleon~\cite{SK}.
The latter became widely appreciated when Witten~\cite{EW} linked the Skyrme 
model to the large $N_c$ limit of QCD. Chiral bag models started with 
ref.~\cite{Cho} but further significant development stalled     
when it was recognized that their failure to treat quark and hadron boosts 
adequately along with the violation of translation invariance is rather  
difficult to correct systematically. Since dynamical quarks, and constituent 
quarks as their low momentum limit, became more widely accepted as appropriate 
degrees of freedom with growing support from NJL models, chiral quark models   
came to dominate the literature.~\cite{WS}    
\par
Chiral fluctuations 
$q_{\uparrow,\downarrow}\rightarrow q_{\downarrow,\uparrow}+(q\bar q')_0$ of 
quarks into pseudoscalar mesons, $(q\bar q')_0$, of the SU(3) flavor octet of 
$0^-$ Goldstone bosons, were first applied to the spin problem of the proton in 
ref.~\cite{EHQ}. It was shown that chiral dynamics can help one understand not 
only the reduction of the proton spin carried by the valence quarks from 
$\Delta \Sigma =1$ in the NQM to the experimental value of about $1/3$, but 
also the reduction of the axial vector coupling constant $g_A^{(3)}$ from the 
NQM value 5/3 to about 5/4. In addition, the violation of the Gottfried sum 
rule~\cite{Go} which signals an isospin asymmetric quark sea in the proton 
became plausible. Here we wish to study the effects of SU(3) breaking which 
are needed to explain the remaining discrepancies of the spin and quark sea 
observables with the data.    
\par
Subsequently the analysis was extended to the $\eta'$ meson~\cite{CL} although 
it is generally not regarded as a Goldstone boson. A singlet pseudoscalar 
coupling constant that differs from that of the octet was shown to cause the 
quark sea to become more flavor asymmetric. Amongst the pseudoscalar mesons 
the $\eta'$ is the heaviest. The Noether current of the $U_A(1)$ symmetry is 
the singlet axial vector current whose divergence contains the $U_A(1)$ 
anomaly. Despite the spontaneous breakdown of the $U_L(1)\times U_R(1)$ 
symmetry, no corresponding Goldstone boson seems to arise because of instanton 
configurations with integral topological charge. Thus the properties of the 
$\eta'$ meson differ significantly from those of other Goldstone bosons such 
as the pions, kaons and the $\eta $ meson. Nonetheless, for the sake of 
comparing with~\cite{CL} in Sect. III we include also broken U(3) flavor 
results on the quark spin fractions with the $\eta'$ meson. Let us now turn to 
the $\eta $ meson case and its problems.    
\par
The $\eta$ meson arises as the octet Goldstone boson when 
the chiral $SU(3)_L\times SU(3)_R$ symmetry is spontaneously broken. 
Predictions from  PCAC are not in good agreement with experiments, e. g. its 
octet Goldberger-Treiman relation is violated because it predicts a fairly 
large $\eta $NN coupling constant which disagrees with the much smaller value 
extracted from analyses of both $p\bar p$ collisions~\cite{GK} and recent 
precision data from MAMI~\cite{TBK} on $\eta $ photoproduction off the proton 
at threshold. Corrections from chiral perturbation theory are of order 30\% 
and therefore much too small to help one understand the problem of the 
suppressed $\eta $NN coupling better~\cite{SW}.    
   
\par
The paper is organized as follows. In Sect. II we describe the formalism of 
SU(3) breaking on the quark spin fractions and the quark sea contents of the 
proton. In Sect.III we numerically evaluate the quark spin distributions 
by considering the SU(3) breaking effect as brought about by the mass 
splitting between the up/down and the strange quarks. The paper concludes 
with a brief summary in Sect. IV.    

\section{SU(3) Breaking}
If the spontaneous chiral symmetry breakdown in the infrared regime of QCD is 
governed by chiral $SU(3)_L \times SU(3)_R$ 
transformations then the effective interaction between the octet of Goldstone 
boson fields $\Phi_i$ and quarks is a flavor scalar given by  
\begin{eqnarray}
{\cal L}_{int}=-{g_A\over 2f_\pi }\sum_{i=1}^{8}\bar q \partial_\mu 
               \gamma^\mu \gamma_5 \lambda_i \Phi_i q.         
\label{lint}
\end{eqnarray}
This interaction will flip the polarization of the quark: 
$q_{\downarrow}\rightarrow q_{\uparrow}+GB$, etc. 
Here $\lambda_i$, $(i=1,2,...,8)$ are the Gell-Mann SU(3) flavor matrices, 
and $g_A$ is the dimensionless axial vector quark coupling constant that is 
taken to be 1, while    
\begin{equation} 
g_A^{(3)}=\Delta u -\Delta d=\Delta_3={\cal F}+{\cal D}
         =(G_A/G_V)_{n\rightarrow p}, 
\label{gan}
\end{equation}
is the isotriplet axial vector coupling constant of the weak decay of the 
neutron, and $\Delta u$, $\Delta d$ and $\Delta s$ stand for the fraction of 
proton spin carried by the u, d and s quarks, respectively. They are defined 
by the following matrix elements of the axial vector currents for the nucleon 
state   
\begin{eqnarray}
\langle N|\bar q\gamma_\mu \gamma_5 {\lambda^3\over 2} q|N\rangle
= g_A^{(3)} \bar U_N\gamma_\mu \gamma_5 {\tau^3\over 2} U_N\, ,\qquad \qquad
&& g_A^{(3)} = \Delta u -\Delta d\, ,\\ 
\label{gA}
\langle N| \bar q \gamma_\mu \gamma_5 {\lambda^8\over 2} q |N \rangle
= {g_A^{(8)}\over 2} \bar U_N \gamma_\mu  \gamma_5 U_N\, ,\qquad\qquad
&& g_A^{(8)} ={1\over \sqrt{3}} (\Delta u +\Delta d -2\Delta s)\, ,\\
\label{gA_8}
\langle N|\bar q\gamma_\mu \gamma_5 {\lambda^0\over 2} q|N\rangle
= {g_A^{(0)}\over 2} \bar U_N\gamma_\mu \gamma_5 U_N\, ,\qquad\qquad
&& g_A^{(0)} = \sqrt{{2\over 3}} (\Delta u +\Delta d + \Delta s)\, ,\\ 
\label{gA_0}
\langle N| \bar s \gamma_\mu \gamma_5 s|N\rangle =
\Delta s\ \bar U_N \gamma_\mu\gamma_5 U_N\, .\qquad\qquad 
\label{G1_s}
\end{eqnarray}
Here $U_N$ is the Dirac spinor of the nucleon and $g_A^{(i)}$ for $i=3,8,0$  
are the nucleon's isovector, hypercharge and singlet  
axial vector couplings, respectively.
It is also common to define the hypercharge spin fraction $\Delta_8$ and the 
total spin $\Delta \Sigma$ as  
\begin{equation}
\Delta_8=\Delta u +\Delta d -2\Delta s=3{\cal F}-{\cal D}, \qquad 
\Delta \Sigma=\Delta u +\Delta d +\Delta s.  
\label{deli}
\end{equation}
\par
The SU(3) symmetric chiral quark model~\cite{EHQ,CL} that invokes 
Goldstone boson (off-mass-shell space-like) fluctuations of constituent 
valence quarks inside hadrons explains several, but not all, spin and sea 
quark observables of the proton. Clearly, the data~\cite{e143,SMC} call for 
SU(3) breaking because some of the spin fractions such as  
$\Delta_3/\Delta_8=$5/3 and the weak axial vector coupling constant of the 
nucleon, $g_A^{(3)}={\cal F}+{\cal D}=$0.85~\cite{EHQ} and 1.12~\cite{CL}, 
respectively, still disagree with experiments in the SU(3) symmetric case. 
The success of hadronic mass relations suggests that a chiral interaction 
which breaks the SU(3) flavor symmetry also be governed by $\lambda_8$, as it 
is expected to originate from the mass difference between the strange and up 
and down quarks (and the corresponding mass differences of the Goldstone 
bosons).  
\par
Writing only the flavor dependence of these interactions we therefore extend 
the SU(3) symmetric Eq.~\ref{lint} 
to 
\begin{eqnarray}   
L_{int}= {g_8\over \sqrt 2} \sum_{i=1}^{8} \bar q (1+\epsilon \lambda_8) 
      \lambda_i \Phi_i q ,
\label{fint}
\end{eqnarray}
\begin{eqnarray}
 {1\over \sqrt 2}\sum_{i=1}^{8}\lambda_i \Phi_i =  \left( \begin{array}{c}  
{ 1\over \sqrt 2} \pi^0 + { 1\over \sqrt 6} \eta \qquad \pi^+ \qquad  K^+ \cr
  \pi^- \quad -{ 1\over \sqrt 2}\pi^0+{ 1\over \sqrt 6}\eta \quad K^0 \cr
   K^- \qquad \bar K^0 \qquad  -{2\over \sqrt 6}\eta  
\end{array}\right) . 
\label{flama}  
\end{eqnarray}
Here $g_8^2:=a \sim f_{\pi NN}^2/4\pi \approx 0.08$ where 
$f_{\pi NN}:=g_{\pi NN} m_\pi/2m_N$ denotes the pseudovector $\pi$N coupling 
constant and $g_{\pi NN}$ the pseudoscalar one. The latter can be related to 
Eq.~\ref{lint} via the pion's Goldberger-Treiman relation $g_{\pi NN}/m_N=
g_A^{(3)}/f_\pi $. Despite the nonperturbative nature of the chiral 
symmetry breakdown the interaction between quarks and Goldstone bosons is 
small enough for a perturbative expansion in $g_8$ to apply. Note also that 
$\epsilon$ is the SU(3) breaking parameter which is expected to satisfy 
$|\epsilon|<1$ in line with the small constituent quark mass ratio 
$m_q/m_s\approx$ 0.5 to 0.6.   

\par
{}From Eq.~\ref{fint} the following transition probabilities 
$P(u_{\uparrow} \rightarrow \pi^+ + d_{\downarrow})$,...
for chiral fluctuations of quarks can be organized as coefficients in the 
symbolic reactions:
\newpage 
\begin{eqnarray} u_{\uparrow} \rightarrow a(1+{\epsilon \over \sqrt{3}})^2 
   (\pi^+ + d_{\downarrow})
   +a(1+{\epsilon \over \sqrt{3}})^2 {1\over 6}(\eta +u_{\downarrow})
   +a(1+{\epsilon \over \sqrt{3}})^2 {1\over 2}(\pi^0 + u_{\downarrow})\cr
   +a(1-{2\epsilon \over \sqrt{3}})^2 (K^+ + s_{\downarrow}),\cr
 d_{\uparrow} \rightarrow a(-1-{\epsilon \over \sqrt{3}})^2 (\pi^- 
   + u_{\downarrow})
   +a(1+{\epsilon \over \sqrt{3}})^2 {1\over 6}(\eta +d_{\downarrow})
   +a(-1-{\epsilon \over \sqrt{3}})^2 {1\over 2}(\pi^0 + d_{\downarrow})\cr
   +a(1-{2\epsilon \over \sqrt{3}})^2 (K^0 + s_{\downarrow}),\cr
 s_{\uparrow} \rightarrow a(1-{2\epsilon \over \sqrt{3}})^2 {2\over 3}(\eta 
   +s_{\downarrow})+a(-1+{2\epsilon \over \sqrt{3}})^2 (K^- + 
    u_{\downarrow}) +a(-1+{2\epsilon \over \sqrt{3}})^2 (\bar K^0 +
    d_{\downarrow}),\cr  
\label{fluc}
\end{eqnarray}
and similar ones for the other quark polarization.  
The Goldstone bosons have the usual quark composition, viz.  
\begin{eqnarray}
|\pi^0\rangle = {1\over \sqrt{2}}(\bar u u- \bar d d), \qquad
|\eta \rangle = {1\over \sqrt{6}}(\bar u u+ \bar d d -2 \bar s s),\qquad
|K^+ \rangle = u \bar s, \qquad etc. 
\label{mes}
\end{eqnarray}  
\par
{}From the u and d quark lines in Eq.~\ref{fluc} the total meson emission 
probability P of the proton is given to first 
order in the Goldstone fluctuations by 
\begin{eqnarray}
P = a[{5\over 3}(1+{\epsilon \over \sqrt{3}})^2  
    +(1-{2\epsilon \over \sqrt{3}})^2]. 
\label{prob}
\end{eqnarray}  

\par
The polarized quark probabilities may now be read off the proton  
composition expression~\cite{EHQ}  
\begin{eqnarray}
(1-P)({5\over 3} u_{\uparrow} + {1\over 3} u_{\downarrow} 
 +{1\over 3} d_{\uparrow} + {2\over 3} d_{\downarrow})
 + {5\over 3} P(u_{\uparrow}) + {1\over 3} P(u_{\downarrow}) 
 + {1\over 3} P(d_{\uparrow}) + {2\over 3} P(d_{\downarrow}).
\label{qp}
\end{eqnarray}
Since the {\bf antiquarks from Goldstone bosons are unpolarized} we use 
$\bar u_\uparrow = \bar u_\downarrow$ in the spin fractions  
$\Delta u = u_\uparrow -u_\downarrow +\bar u_\uparrow -\bar u_\downarrow$, etc 
and $\Delta s=\Delta s_{sea}$, $\Delta \bar u=\Delta \bar d=\Delta \bar s =0$.  
Moreover, the valence quark fractions are (see the NQM values in Table 1) 
$\Delta u_v=4/3,\quad  \Delta d_v=-1/3, \quad \Delta s_v=0$. Altogether then 
Eq.~\ref{qp}, in conjunction with the probabilities displayed 
in Eq.~\ref{fluc}, yields the following spin fractions  
\begin{eqnarray}
\Delta u=u_{\uparrow}-u_{\downarrow}={4\over 3}(1-P)
 -{5\over 9}a(1+{\epsilon\over \sqrt{3}})^2,
\label{del1}
\end{eqnarray}
\begin{eqnarray}
\Delta d=-{1\over 3}(1-P)-{10\over 9}a(1+{\epsilon \over \sqrt{3}})^2, 
\qquad 
\label{del2}
\end{eqnarray}
\begin{equation}
\Delta s=-a(1-{2\epsilon\over \sqrt{3}})^2.  \qquad
\label{del3}
\end{equation}

\par
If the antisymmetrization of the up and down sea quarks with the valence quarks 
is ignored we may assume that $u_v=2,\ d_v=1,\ s_v=0$ and 
$u_{sea}=\bar u$,\ etc so that 
\begin{equation}
u=2+\bar u, \qquad d=1+\bar d, \qquad s=\bar s, 
\label{us}
\end{equation}
reflecting equal sea quark and antiquark numbers. From Eqs.~\ref{fluc},
\ref{mes},\ref{qp} we now obtain the antiquark fractions   
\begin{equation}
\bar u=2a(1+{\epsilon\over \sqrt{3}})^2,
\label{anti1}
\end{equation}

\begin{equation}
\bar d={8\over 3}a(1+{\epsilon\over \sqrt{3}})^2, 
\label{anti2}
\end{equation}
\begin{eqnarray}
\bar s=3a(1-{2\epsilon\over \sqrt{3}})^2
       +3a[-{1\over 3}(1+{\epsilon\over \sqrt{3}})]^2.
\label{anti3}
\end{eqnarray}
{}From Eqs.~\ref{anti1},\ref{anti2},\ref{anti3} it is obvious that the sea 
violates the SU(3) flavor and isospin symmetries. We also see that for the 
broken SU(3) case $\bar u/\bar d=$3/4 is still the same as in the 
SU(3) symmetric case $\epsilon =$0.~\cite{EHQ} 

\par  
The Gottfried sum rule
\begin{equation} 
I_G=\int_{0}^{1} {dx\over x}[F_2^p(x)-F_2^n(x)]={1\over 3}+{2\over 3}
    (\bar u - \bar d),
\label{GSR}
\end{equation}
where x is the Bjorken scaling variable and $F_2^{p,n}(x)$ the unpolarized 
nucleon structure functions, measures the isospin asymmetry, $\bar u-\bar d$, 
of the antiquarks. The antiquark flavor fractions are generally defined as  
\begin{eqnarray}
f_q=(q+\bar q)/\sum_{q=u,d,s}(q+\bar q), \qquad for \qquad q=u,d,s,
\label{fq}
\end{eqnarray}
\begin{eqnarray} 
f_3=f_u-f_d, \qquad f_8=f_u+f_d-2f_s, \qquad 
f_s=2\bar s/[3+2(\bar u + \bar d + \bar s)].  
\label{fs}
\end{eqnarray}

\section{Numerical Results}
\par 
When SU(3) breaking 
is included that is consistent with the higher mass of the strange quark 
compared to the common up, down quark mass and is governed by the hypercharge 
generator $\lambda_8$, then nearly all observables agree with the data.    
\par
In fact, with SU(3) breaking that is characterized by the parameter 
$\epsilon$ defined in Eq.~\ref{fint}, and the parameter values $a=$0.12 and 
$\epsilon =$0.2 (see the 4th column in Table 1) the spin fraction ratio 
$\Delta_3/\Delta_8$ increases from the value 5/3 of the SU(3) symmetric case 
for $\epsilon=$0~\cite{EHQ,CL} and the NQM to 2.12, 
which is much closer to the experimental value $2.09\pm 0.13$~\cite{e143}. 
The situation is 
similar for the fraction $f_3/f_8$~\footnote{The experimental value for the 
antiquark fraction $f_3/f_8$ is obtained from the measured octet baryon 
masses.} decreasing from the value 1/3 for 
$\epsilon=0$ (and the NQM, cf. Table 1) to 0.24 for $\epsilon=$0.2 close to 
the experimental value 0.23$\pm$0.05. 
A significant defect seems to remain despite SU(3) 
breaking in so far as the axial vector nucleon coupling constant 
$g_A^{(3)}={\cal F}+{\cal D}=1.217$ for $\epsilon=0.2$ is below the 
experimental value $1.2573\pm 0.0028$~\cite{PDG}. In view of missing 
relativistic effects, which are known to drive this quantity even lower, 
this discrepancy and possibly $\bar u/\bar d=$3/4 are the only ones remaining 
in the broken SU(3) case. Overall, SU(3) breaking leads to markedly improved 
results for the $SU(3)_L \times SU(3)_R$ chiral quark model.      
\par
Another description of quark spin fractions, where $\epsilon_{SMW}$ 
parametrizes the suppression of kaon transitions only, has recently been given 
in~\cite{SMW}. Upon comparing our $\Delta s=-a(1-{2\epsilon\over \sqrt{3}})^2$ 
from Eq.~\ref{del3} with their $\Delta s=-a\epsilon_{SMW}$ we obtain 
$\epsilon=(1-\sqrt{\epsilon_{SMW}}){\sqrt{3}\over 2}$, and using their fit 
values $\epsilon_{SMW}\approx 0.5-0.6$ we find the estimates 
\begin{equation}
0.195 \approx (1-\sqrt{0.6}){ \sqrt{3}\over 2} < \epsilon < 
 (1-\sqrt{0.5}){ \sqrt{3}\over 2}\approx 0.25 ,
\label{eps}
\end{equation}
which are in reasonable agreement with the value, $\epsilon=$0.2, that we 
establish in Table 1.   
\par
Let us also compare with the case where the singlet $\eta '$ meson is included 
in chiral meson-quark interactions with a relative coupling constant $\zeta $ 
that differs from that of the octet.~\cite{CL}. Despite varying the additional 
parameter $\zeta $, the fit in the fifth column of Table 1 for the case with 
SU(3) breaking hardly improves the case without $\eta '$ meson in the 
4th column, except possibly for $\bar u/\bar d$ decreasing from 3/4 to 0.686. 
In particular, the inclusion of the $\eta '$ meson does not    
resolve the discrepancy with the nucleon axial vector coupling 
constant. Since the relative $\eta '$ coupling, $\zeta =$-0.3, turns out to be 
much smaller than in the SU(3) symmetric case, where $\zeta =$-1.2, 
the $\eta' $ meson becomes almost negligible in the broken SU(3) case.   
\par 
Can the remaining discrepancies in the broken SU(3) case be better 
understood? As we mentioned in the introduction, the Goldberger-Treiman 
relation of the $\eta $ meson is in conflict with experiments which can be 
avoided if it couples only to the strange, but not the u and d, quarks. If we 
assume that to be the case in the last column of Table 1, we see that both 
remaining discrepancies become less pronounced. 
In fact, $g_A^{(3)}={\cal F}+{\cal D}$ increases to 1.335 (and this high value 
is likely to be beneficial when relativistic effects are included) and 
$\bar u/\bar d$ decreases from 3/4 to 7/11, while the other spin and flavor 
fractions change, but not by much.

\section{Summary and Conclusion}
\par
We have seen that in the broken SU(3) case nearly all of the nucleon's spin 
observables are reproduced by the $SU(3)_L\times SU(3)_R$ chiral quark model, 
where the $\eta $ meson is the conventional octet Goldstone boson. The 
nucleon's axial vector coupling constant $g_A^{(3)}$ may not be large enough, 
though, because relativistic effects are not included here which are known to 
drive this quantity to lower values. Including the $\eta '$ meson in the 
chiral dynamics does not seem to help one much to understand better the proton 
spin problem.    
\par
When the $\eta$ meson is taken to couple only to the strange, but not the u 
and d, quarks in the chiral quark model,  
then $g_A^{(3)}={\cal F}+{\cal D}$ increases and the fit improves for 
$\bar u/\bar d$ as well.    
\par
The remarkable improvement in the spin and flavor fractions from SU(3) 
breaking shows that such chiral quark models provide a sound phenomenological 
framework for understanding the spin problem of the proton. 

\section{Acknowledgement}
The work of M.K. was supported by Deutsche Forschungsgemeinschaft (SFB 201).
The work of X.S. was supported in part by the U.S. Department of Energy.  

\vfill\eject

{\bf Table 1}\qquad Quark Spin and Sea Observables of the Proton
$$
\offinterlineskip \tabskip=0pt 
\vbox{ 
\halign to 1.0\hsize 
   {\strut
  \vrule#                          
   \tabskip=0pt plus 30pt
 & \hfil #  \hfil                  
 & \vrule#                         
 & \hfil #  \hfil                  
 & \vrule#                         
 & \hfil #  \hfil                  
 & \vrule#                         
 & \hfil #  \hfil                  
 & \vrule#                         
 & \hfil #  \hfil                  
 & \vrule#                         
 & \hfil #  \hfil                  
   \tabskip=0pt                    %
 & \vrule#                         
  \cr                             
\noalign{\hrule}
&\quad Observable &&\quad Data &&\quad NQM && $a=$0.12 && $a=$0.12 && $a=$0.16 
&\cr
& && &&  &&   && $\zeta =$-0.3  && $\eta $\ mod. &\cr
& &&\quad Ref.\cite{e143}&& && $\epsilon =$ 0.2  && $\epsilon =$ 0.2 
 && $\epsilon =$ 0.2 &\cr
\noalign{\hrule}
& $\Delta u$ && 0.84$\pm$0.05 && 4/3 &&  0.824 && 0.81 &&0.87 &\cr
& $\Delta d$ &&-0.43$\pm$0.05 && -1/3 &&-0.39 &&-0.39 &&-0.47 &\cr
& $\Delta s$ &&-0.08$\pm$0.05&&0&& -0.07 &&-0.07 &&-0.095 &\cr
& $\Delta \Sigma $ && 0.30$\pm$0.06&&1&& 0.36 && 0.35 && 0.31 &\cr
& $\Delta_3/\Delta_8$ &&2.09$\pm$0.13&&5/3&& 2.12 && 2.13 && 2.255 &\cr
& ${\cal F}+{\cal D}$  && 1.2573$\pm$0.0028 && 5/3 && 1.217 &&
1.205 && 1.335 &\cr
& ${\cal F}/{\cal D}$ && 0.575$\pm$0.016 && 2/3 && 0.58 && 0.58 
&& 0.565 &\cr
& $\bar u/\bar d$ &&0.51$\pm$0.09 && 1 &&0.75 && 0.686 && 0.636 &\cr
& $f_3/f_8$  &&0.23$\pm$0.05 && 1/3 && 0.24 && 0.235 &&0.165 &\cr
& $I_G$      &&0.235$\pm$0.026 && 1/3 && 0.27 && 0.25 && 0.20 &\cr
\noalign{\hrule}
}}$$

\vfill\eject

\end{document}